      [1999/12/01 v1.4c Il Nuovo Cimento]
\documentclass{cimento}

\usepackage{graphicx}  % got figures? uncomment this
\title{Azimuthal distributions of pions inside jets at RHIC}
\author{U.~D'Alesio\from{ins:uni}\from{ins:infn}\ETC,
F.~Murgia\from{ins:infn}
        \atque
C.~Pisano\from{ins:uni}\from{ins:infn}\thanks{Speaker. Talk given at 3rd Workshop on the QCD structure of the nucleon (QCD-N'12), Bilbao (Spain), 22-26 October 2012.}}
\instlist{\inst{ins:uni} Dipartimento di Fisica, Universit\`a di Cagliari, Cittadella Universitaria, Monserrato, Italy
  \inst{ins:infn} Istituto Nazionale di Fisica Nucleare, Sezione di Cagliari, C.P.\ 170, Monserrato, Italy}
\PACSes{%\PACSit{00.00}{By the way, which PACS is it, the 00.00? GOK.}
\PACSit{13.85.Ni}{Inclusive production with identified hadrons}\PACSit{13.88.+e}{Polarization in interactions and scattering}}
\begin{document}

\maketitle

\begin{abstract}
We evaluate the azimuthal asymmetries for the distributions of leading pions 
inside a jet, produced in high-energy proton-proton collisions, in kinematic 
configurations currently under active investigation at RHIC. 
Adopting a transverse momentum dependent approach, which 
assumes the validity of a perturbative QCD factorization scheme and takes 
into account all the spin and intrinsic parton motion effects, we show 
how the main  mechanisms underlying these asymmetries, namely the Sivers and
the Collins effects, can be disentangled. Furthermore, we consider the impact 
of color-gauge invariant initial and final state interactions and suggest a
method for testing the universality properties of the Sivers function for 
quarks.
\end{abstract}

\section{Introduction}

We study the process $p^{\uparrow}p\to{\rm jet} \,\pi+X$, where one of 
the protons is transversely polarized and the jet has a large 
transverse momentum. Specifically, we calculate the 
azimuthal asymmetries in the distribution of leading pions around the jet axis
within the so-called transverse momentum dependent framework, 
which takes into account all the possible spin and intrinsic parton motion 
effects, assuming  the validity of perturbative QCD factorization. 
Within this approach, the leading-twist azimuthal asymmetries are expressed 
in terms of convolutions of different 
transverse momentum dependent distribution and fragmentation 
functions (TMDs) \cite{D'Alesio:2010am}. We focus on the ones which 
are the most relevant from a 
phenomenological point of view, namely the Sivers distribution function and 
the Collins 
fragmentation function.  By defining appropriate moments of 
the asymmetries, it is possible to separate the effects due to these
two different TMDs, in strong analogy with the case of 
semi-inclusive  deep inelastic scattering (SIDIS). This will also help
in clarifying the role played by the Sivers and the Collins 
mechanisms in the sizable single-spin asymmetries observed at RHIC for 
inclusive pion production, where these two effects cannot be 
isolated.

\section{ Theoretical framework }

In the hadronic center-of-mass, or in any other frame connected to it by a
boost along the direction of the two colliding protons, the single-transverse 
polarized cross section for the process  $p^{\uparrow}p\to{\rm jet} \,\pi+X$,
at leading order in pQCD, can be written as \cite{D'Alesio:2010am}
\begin{eqnarray}
2{\rm d}\sigma(\phi_{S},\phi_\pi^H) &\sim & {\rm d}\sigma_0
\, +\, {\rm d}\Delta\sigma_0\sin\phi_{S} \, +\, 
{\rm d}\sigma_1\cos\phi_\pi^H \, +\,  
{\rm d}\sigma_2\cos2\phi_\pi^H \nonumber \\
&& \quad +  \, {\rm d}\Delta\sigma_{1}^{-}\sin(\phi_{S}-\phi_\pi^H)\, +\, 
{\rm d}\Delta\sigma_{1}^{+}\sin(\phi_{S}+\phi_\pi^H)\nonumber \\
&& \quad\quad+\, {\rm d}\Delta\sigma_{2}^{-}\sin(\phi_{S}-2\phi_\pi^H) 
+ \, {\rm d}\Delta\sigma_{2}^{+}\sin(\phi_{S}+2\phi_\pi^H)\,,
\label{d-sig-phi-SA}
\end{eqnarray} 
with  $\phi_S$ being the azimuthal angle of the proton spin vector $S$ 
relative to the jet production plane and $\phi_\pi^H$ being the 
azimuthal angle of the three-momentum of the pion around the jet axis, 
as measured in the helicity frame of the fragmenting 
quark or gluon \cite{D'Alesio:2010am}. 

By means of the azimuthal moments
\begin{equation}
A_N^{W(\phi_{S},\phi_\pi^H)}
=
2\,\frac{\int{\rm d}\phi_{S}{\rm d}\phi_\pi^H\,
W(\phi_{S},\phi_\pi^H)\,[{\rm d}\sigma(\phi_{S},\phi_\pi^H)-
{\rm d}\sigma(\phi_{S}+\pi,\phi_\pi^H)]}
{\int{\rm d}\phi_{S}{\rm d}\phi_\pi^H\,
[{\rm d}\sigma(\phi_{S},\phi_\pi^H)+
{\rm d}\sigma(\phi_{S}+\pi,\phi_\pi^H)]}\,,
\label{gen-mom}
\end{equation} 
where $W(\phi_{S},\phi_\pi^H)$ is one of the circular functions in 
(\ref{d-sig-phi-SA}), it is possible to single out the 
different angular modulations of the cross section. 
An estimate of the upper bounds of all the azimuthal moments has been given 
in  \cite{D'Alesio:2010am} for the ongoing experiments at RHIC. In the 
following we will focus only on those (sizable) asymmetries 
that receive contributions from the Sivers and the Collins functions. These 
TMDs are known and their parameterizations have been determined from 
independent fits to $e^+e^-$ and  SIDIS data.

\begin{figure}[t]
\begin{center}
 \includegraphics[angle=0,width=0.4\textwidth]{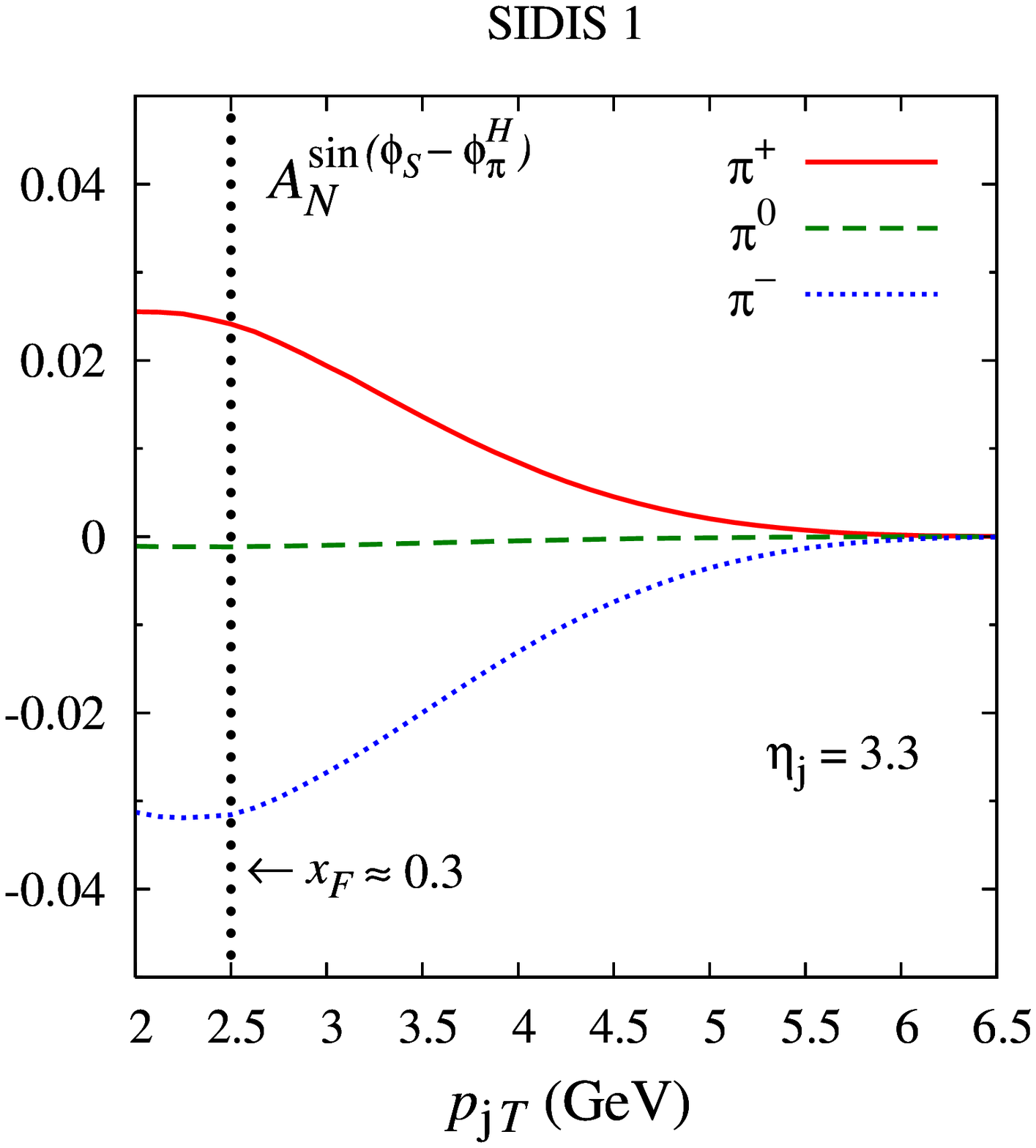}
 \includegraphics[angle=0,width=0.4\textwidth]{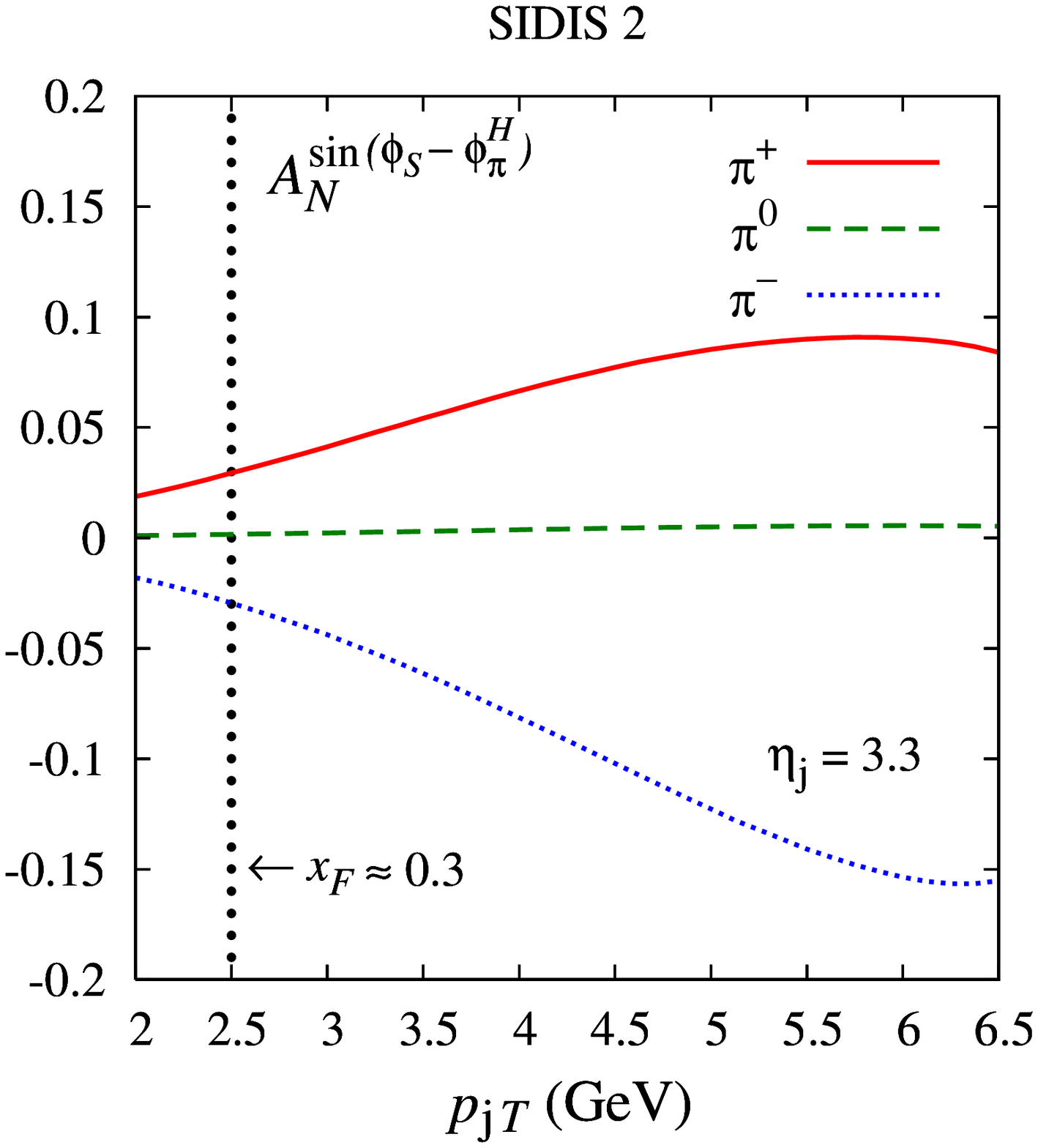}
 \caption{Estimate of the Collins asymmetry 
$A_N^{\sin(\phi_{S}-\phi_\pi^H)}$ in the GPM approach adopting the two sets of parameterizations SIDIS~1 (left panel) and SIDIS~2 (right panel), at $\sqrt{s}=200$ GeV and fixed jet rapidity $\eta_{\rm j} =3.3$, as a function of the transverse
momentum of the jet $p_{{\rm j}T}$.
\label{asy-an-coll-par200} }
\end{center}
\end{figure}
\begin{figure}[t]
\begin{center}
 \includegraphics[angle=0,width=0.35\textwidth]{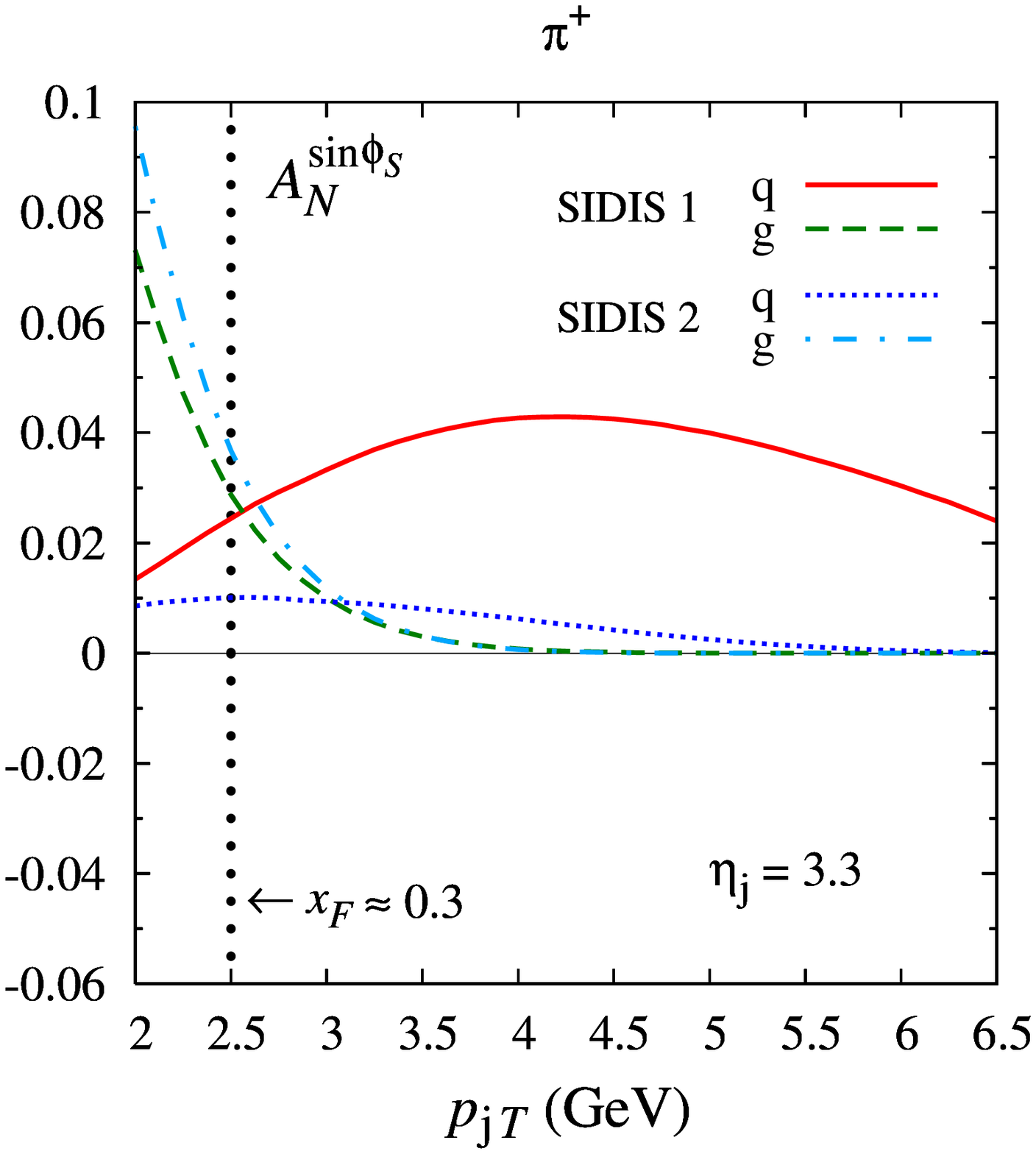}
 \hspace*{-20pt}
 \includegraphics[angle=0,width=0.35\textwidth]{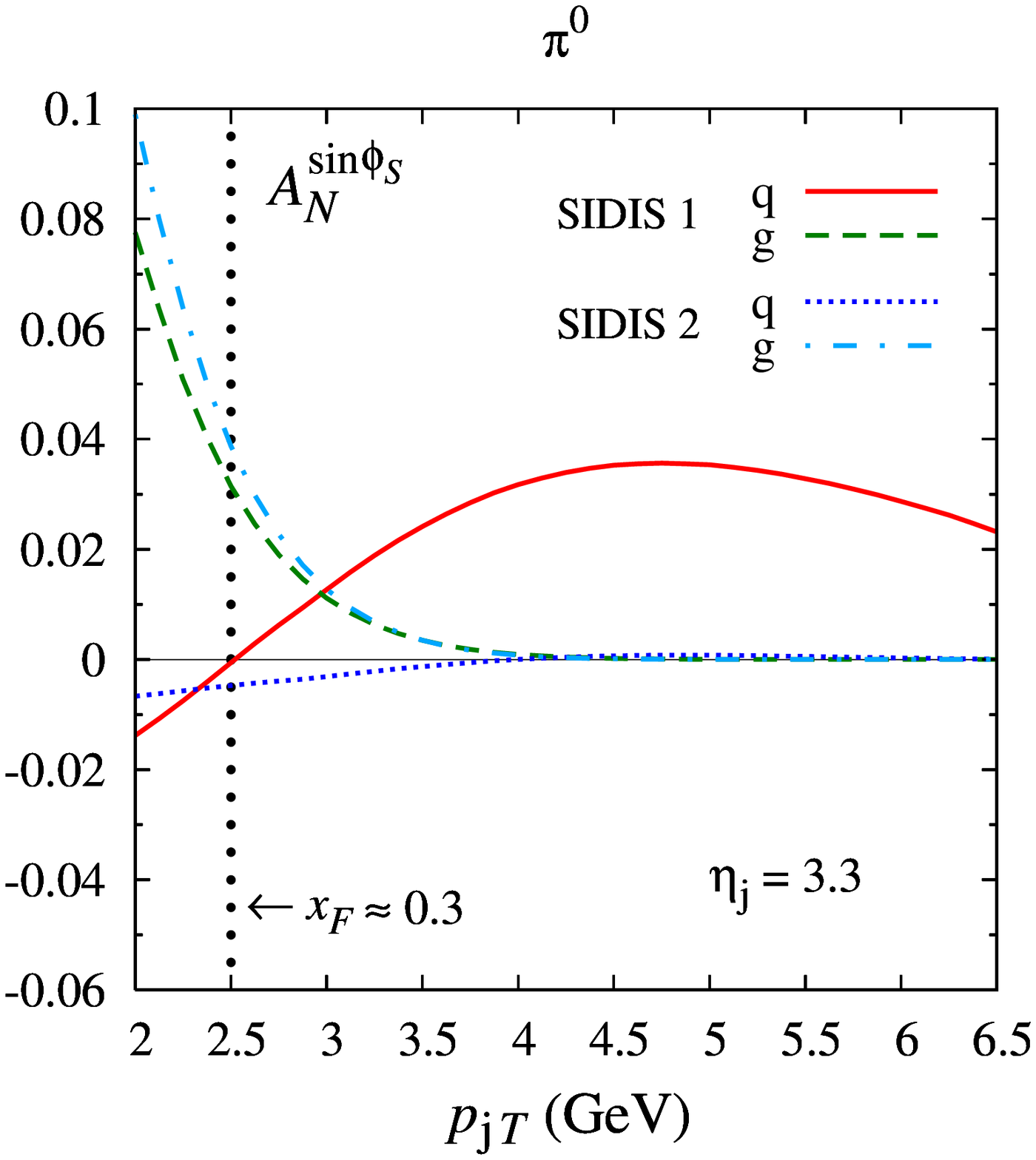}
 \hspace*{-20pt}
 \includegraphics[angle=0,width=0.35\textwidth]{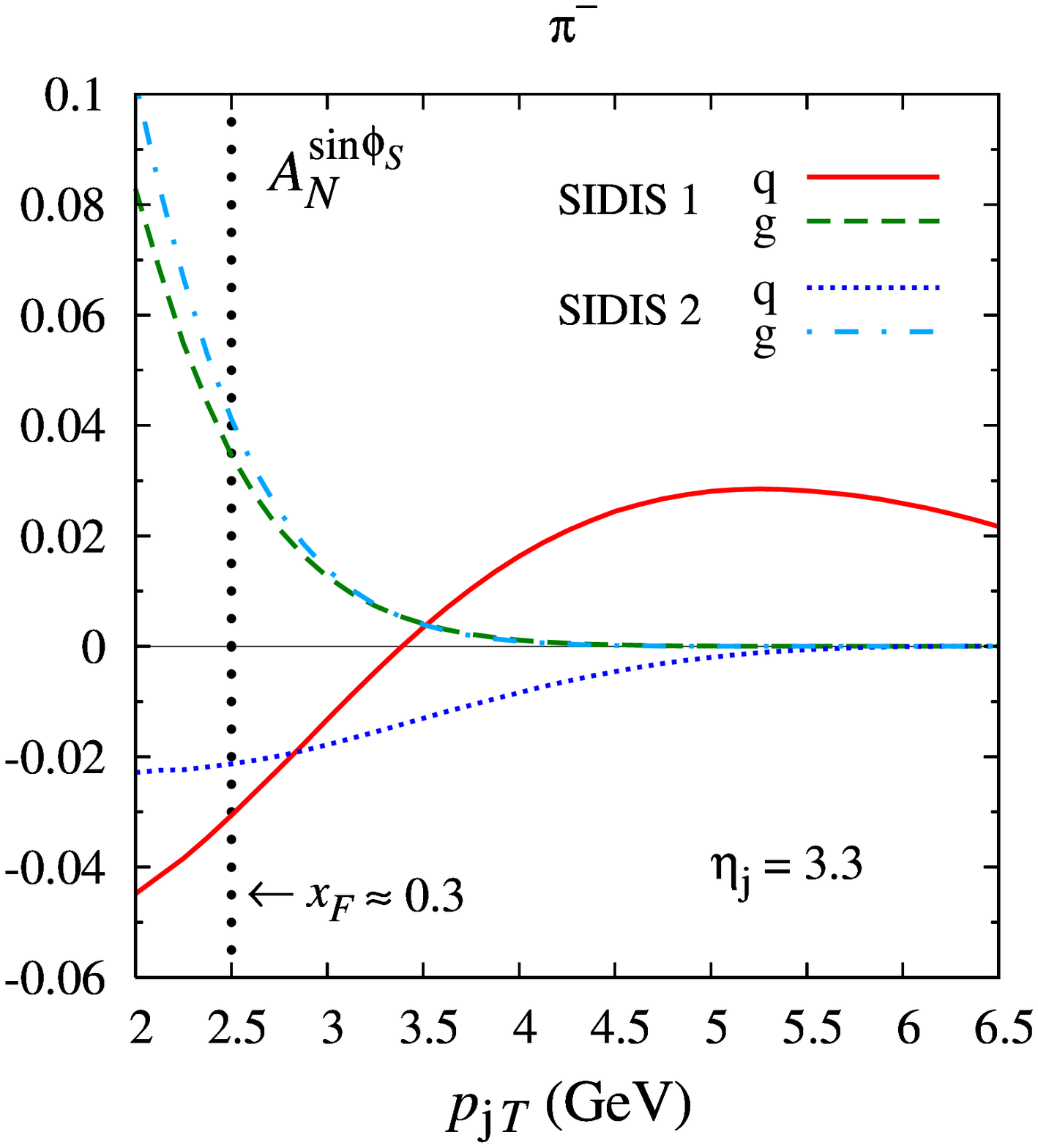}
 \caption{The quark and gluon contributions to the Sivers asymmetry  
$A_N^{\sin\phi_{S}}$ obtained in the GPM approach adopting the two sets of parameterizations SIDIS~1 and SIDIS~2, at $\sqrt{s}=200$ GeV and 
fixed jet rapidity $\eta_{\rm j} =3.3$, as a function of the transverse
momentum of the jet $p_{{\rm j}T}$.
 \label{asy-an-siv-par200} }
\end{center}
\end{figure}

\section{Collins and Sivers asymmetries in the generalized parton model}

In this section the Collins and Sivers asymmetries are evaluated at 
the RHIC energy $\sqrt s = 200$ GeV and at forward jet rapidity, 
within the generalized parton  model (GPM) approach, according to 
which TMDs are considered to be process independent \cite{D'Alesio:2010am}. 
In our analysis, two different sets of parameterizations for the TMDs 
have been used, named SIDIS~1 and SIDIS~2  \cite{D'Alesio:2010am}.   

The Collins asymmetry $A_N^{\sin(\phi_{S}-\phi_\pi^H)}$, presented in Fig.~\ref{asy-an-coll-par200}, is mainly given by a 
convolution of the Collins fragmentation function and the transversity distribution. We note that our prediction of a negligible value of $A_N^{\sin(\phi_{S}-\phi_\pi^H)}$ for neutral pions has been confirmed by preliminary RHIC data \cite{Poljak:2011vu}.
The Sivers asymmetry $A_N^{\sin\phi_{S}}$ is shown 
in Fig.~\ref{asy-an-siv-par200}. Its quark and gluon contributions are 
depicted 
separately, but cannot be  disentangled experimentally.  
In order to provide an estimate of the unknown gluon Sivers function, 
we have taken it positive, saturating an upper bound derived
from the analysis of PHENIX data for central production of neutral pions 
\cite{Anselmino:2006yq}. Recently the STAR collaboration at RHIC reported 
preliminary data on the Sivers asymmetry for neutral pions, which
turns out to be larger than zero \cite{Poljak:2011vu} and compatible with 
our results obtained within the GPM framework.

The vertical dotted lines in Figs.~\ref{asy-an-coll-par200} and \ref{asy-an-siv-par200} delimit the range
$x_F \le 0.3$ in which TMDs are presently constrained by SIDIS data. 
Their extrapolation beyond this region leads to results plagued by large 
uncertainties. Hence a measurement of the proposed observables would  
shed light on the large $x$ behavior of the Sivers and the 
transversity distribution functions.

\section{The Sivers asymmetry in the color gauge invariant parton model}

\begin{figure}[t]
\begin{center}
\includegraphics[angle=0,width=0.35\textwidth]{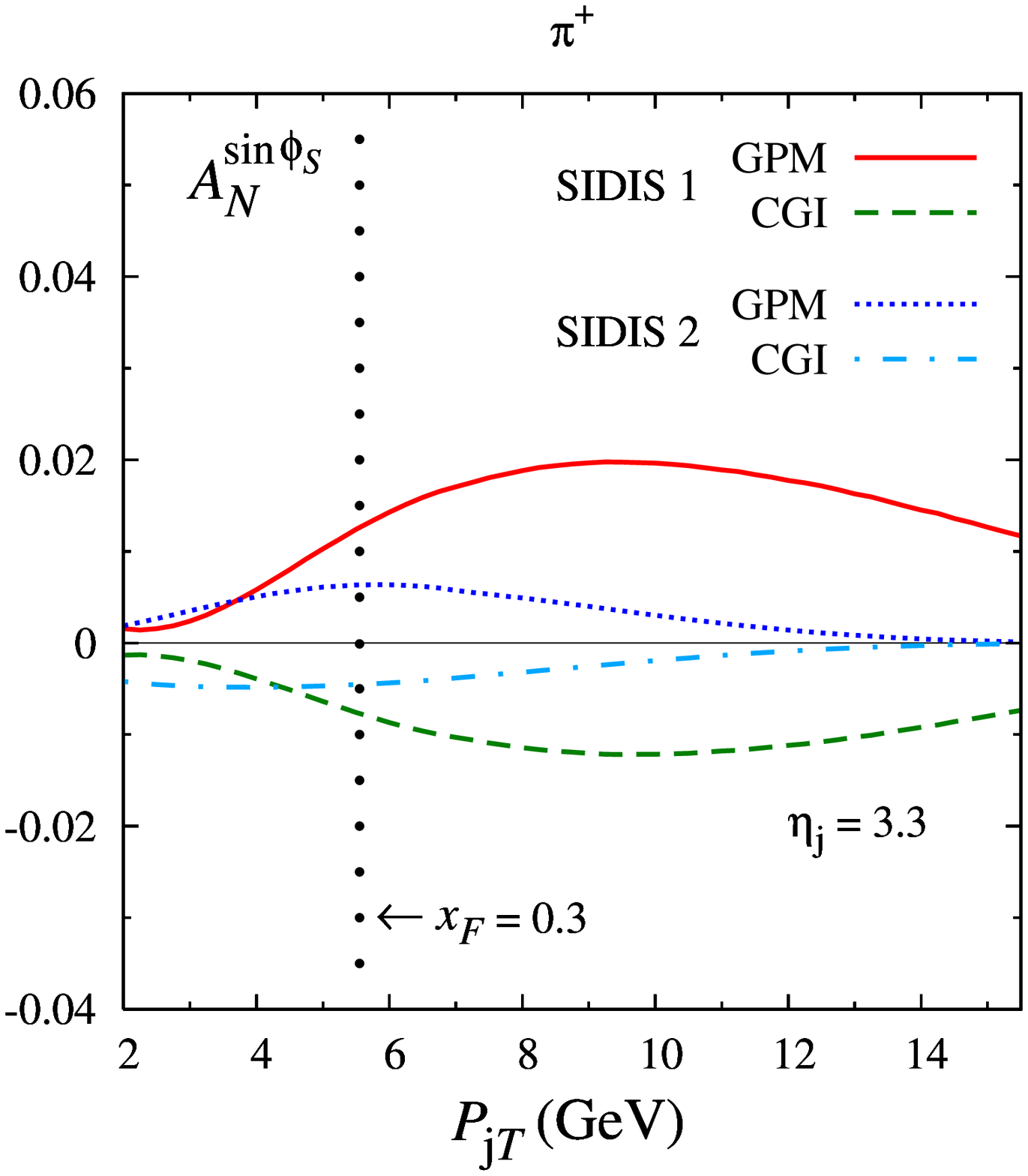}
 \hspace*{-20pt}
 \includegraphics[angle=0,width=0.35\textwidth]{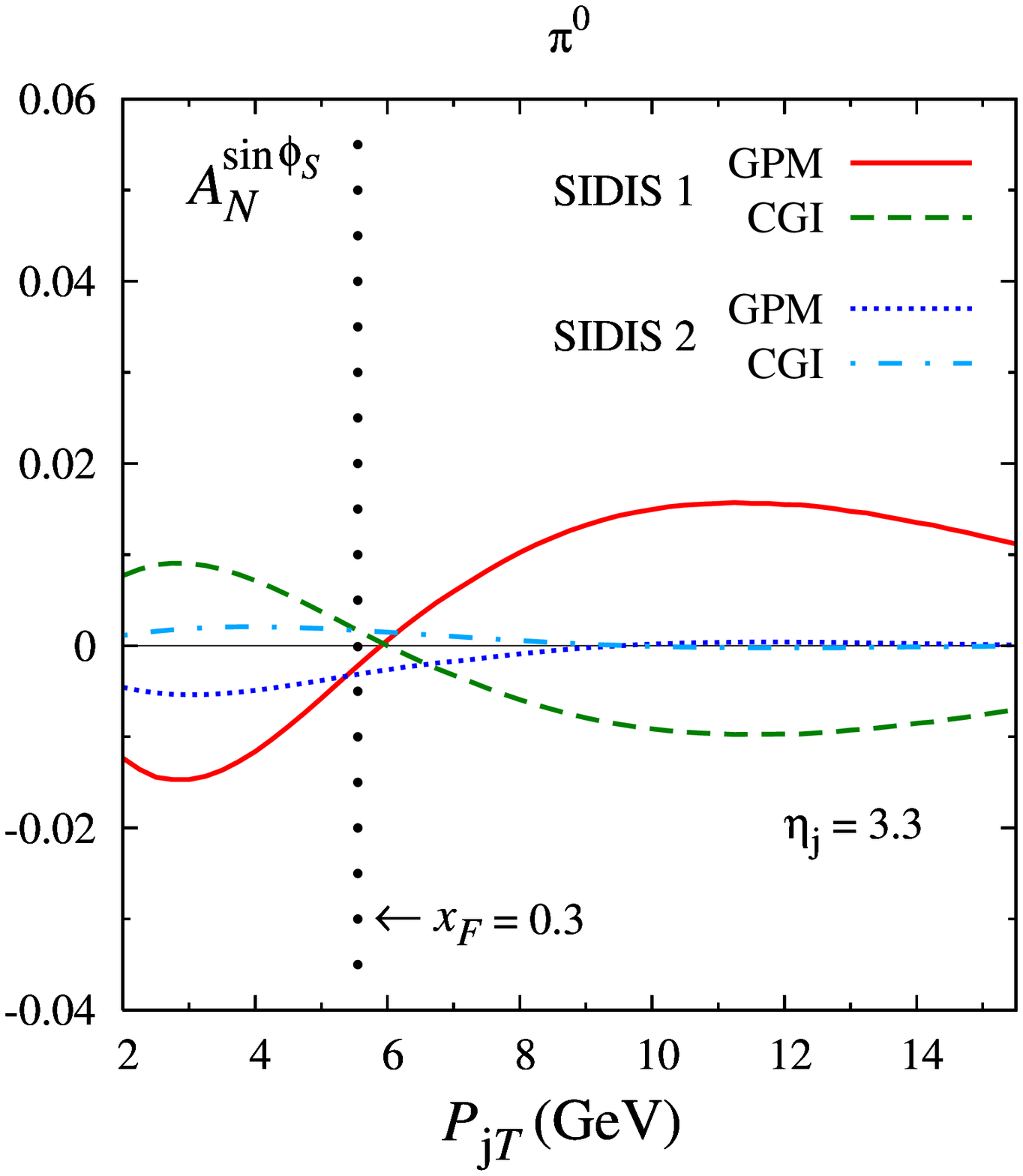}
 \hspace*{-20pt}
 \includegraphics[angle=0,width=0.35\textwidth]{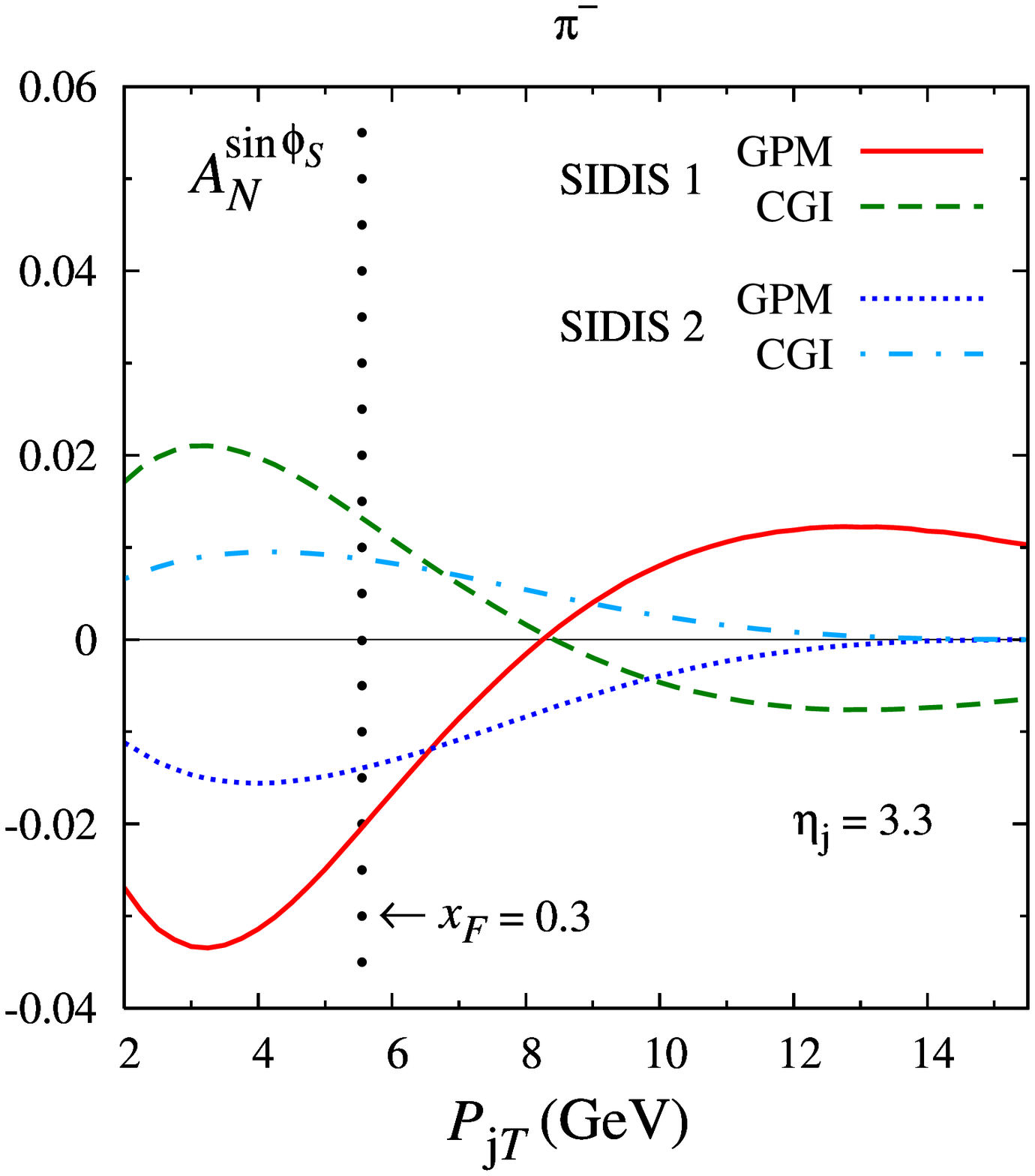}
\caption{The quark contribution to the Sivers asymmetry $A_N^{\sin\phi_{S}}$ calculated  in the GPM and in the CGI GPM approaches adopting two sets of parameterizations, SIDIS~1 and SIDIS~2, at $\sqrt{s}=500$ GeV and jet rapidity 
$\eta_{\rm j} =3.3$, as a function of the jet transverse momentum  $p_{{\rm j}T}$. }
\label{fig3}
\end{center}
\end{figure}

In contrast to the GPM approach adopted in the previous section, in the 
color gauge invariant (CGI) GPM \cite{Gamberg:2010tj,D'Alesio:2011mc} 
TMDs can be process dependent, due to the effects of initial (ISI) and final 
(FSI) state interactions. A fundamental example (still to be confirmed by 
experiments) is provided by the ISI in 
SIDIS and the FSI in the DY processes, which lead to two quark
Sivers functions with an opposite relative sign. For the reaction under study
the quark Sivers function has in general a more involved color structure, 
since both ISI and FSI contribute \cite{D'Alesio:2011mc}. 
However, at forward rapidities  only the $qg\to qg$ channel 
gives a dominant contribution. As a consequence, our predictions for the 
Sivers asymmetries obtained with and without ISI and FSI are
comparable in size but have opposite signs, as depicted 
in Fig.~\ref{fig3} at the RHIC energy $\sqrt{s} = 500$ GeV. 
Therefore the measurement of a sizable asymmetry would verify one of
 the two approaches and test the process dependence of the Sivers function.

Finally, we have also studied single-spin asymmetries for inclusive 
jet production, which are described solely by the Sivers function 
\cite{D'Alesio:2010am,D'Alesio:2011mc}.  Our predictions for 
$A_N^{\sin\phi_{S}}$ turn out to be very similar to 
the ones for jet-neutral pion production, shown in the central panel 
of Fig.~\ref{fig3}. According to preliminary data reported 
by the AnDY Collaboration at RHIC, which have been analyzed very recently
in the different framework of the twist-3 collinear 
formalism \cite{Gamberg:2013kla}, the Sivers asymmetry for 
$p^\uparrow p\to {\rm jet}\,+ X$ is small and positive \cite{Nogach:2012}. 
These results seem to agree with the GPM predictions only for 
$x_F \ge 0.3$ and suggest the need for further studies along these lines, 
aiming  to confirm or disprove the validity of our factorization
hypotesis and to test the universality properties of the different TMDs. 
Currently, a further comparison with experiments is ongoing 
\cite{D'Alesio:2013}, in which the imposed kinematic cuts are as close
 as possible to the ones used at RHIC in the analysis of the azimuthal 
asymmetries for the processes $p^{\uparrow}p\to{\rm jet} \,\pi+X$   and $p^\uparrow p\to {\rm jet}\,+ X$.

\end{document}